\documentclass[3p,times]{elsarticle}

\usepackage{ecrc}

\volume{00}

\firstpage{1}

\journalname{Physics of The Dark Universe }

\runauth{S.Mukherjee et al.}

\usepackage{amssymb}

\begin{document}

\begin{frontmatter}



\title{An accelerated universe with negative equation of state parameter in Inhomogeneous Cosmology with $k$-essence scalar field}


\author[label1,label2]{Somnath Mukherjee, Debashis Gangopadhyay} 
\address[label1]{Department of Physics, Dharsa Mihirlal Khan Institution [H.S], P.O-G.I.P. Colony, Howrah-711112, India, sompresi@gmail.com}
\address[label2]{Department of Physics, School of Natural Sciences, Sister Nivedita University, DG 1/2, Action Area 1 Newtown, kolkata 700156, India, 
 debashis.g@snuniv.ac.in}
\begin{abstract}
We obtain a scaling relation for spherically symmetric k-essence scalar fields 
$\phi(r,t)$ for an inhomogeneous cosmology with the Lemaitre-Tolman- Bondi (LTB) metric. We show that this scaling relation reduces to  the known relation for  a homogeneous cosmology when the LTB metric reduces to the Friedmann-Lemaitre-Robertson-Walker (FLRW) metric under certain identifications of the metric functions. A k-essence lagrangian is set up and the Euler-Lagrangian equations solved assuming $\phi(r,t)=\phi_{1}(r) + \phi_{2}(t)$. The solutions enable the LBT metric functions to be related to the fields. The  LTB inhomogeneous universe exhibits accelerated expansion i.e.cosmic acceleration driven by negative pressure.
\end{abstract}


\end{frontmatter}

\section{Introduction}
\label{intro}
Inhomogeneous cosmological models \cite{Ref1,Ref2,Ref3,Ref24,Ref25,Ref26,Ref27,Ref28,Ref29,Ref30} are those exact solutions of Einstein's equations that contain at least a subclass of non-vacuum, non-static FLRW solutions as a limit. Inhomogeneous models of the universe arise from perturbations of the FLRW metric. A natural question is whether the phenomenon of late time accelerated expansion  persists in such perturbations of the FLRW metric. One relevant spherically symmetric, inhomogeneous metric is the plane Lemaitre-Bondi-Tolman metric given by \cite{Ref3}
\begin{equation}
ds^{2}=-dt^{2}+Y'^{2}dr^{2}+Y^{2}[d{\theta}^{2}+\sin^{2}\theta d{\phi}^{2}]
\end{equation}
where $Y=Y(r,t)$ so that $Y'=\frac{\partial Y}{\partial r}$

The stress-energy tensor is \cite{Ref3}
\begin{equation}
T_{ab}=(\rho+P)u_{a}u_{b}+\left(P-\frac{\Lambda}{8 \pi G}\right)g_{ab}
\end{equation}
where $\rho$ is the energy density,$P$ the pressure and $\Lambda$ the cosmological constant.
However, here $\Lambda$ {\it will not be taken as the source of 
dark energy}. Rather, $k-$essence scalar fields \cite{Ref4,Ref5,Ref6,Ref7,Ref8,Ref9,Ref10,Ref11,Ref12,Ref13,Ref14,Ref15,Ref16,Ref17,Ref18,Ref19,Ref20,Ref21,Ref22,Ref23} will play that role.
Let 
\begin{equation}
 Y(r,t)=f^{\frac{2}{3}}(r,t)
\end{equation}
Then Einstein's equations (with the cosmological constant $\Lambda$) yield 
\begin{equation}
\rho+\Lambda=\frac{4}{3}\frac{\dot{f}\dot{f}'}{ff'}
\end{equation}
 and
\begin{equation}
P-\Lambda=-\frac{4}{3}\frac{\ddot{f}}{f}
\end{equation}
Here we take $8\pi G=1$ and assume 
\begin{equation}
f(r,t)=R(r)T(t)
\end{equation}
Then Einstein's equations $(4)$ and $(5)$ become
\begin{equation}
\rho=\frac{4}{3}\left(\frac{\dot{T}}{T}\right)^{2}-\Lambda
\end{equation}
\begin{equation}
P=-\frac{4}{3}\frac{\ddot{T}}{T}+\Lambda 
\end{equation}

Using $(7)$ and $(8)$ we obtain
\begin{equation}
\dot{\rho}+2\frac{\dot{T}}{T}(\rho+P)=0
\end{equation}
This is the conservation equation for inhomogeneous cosmology, with assumption $(6)$.

\section{Scaling relation for the fields}
\label{sec:1}
The non-canonical Lagrangian for the $k-$essence field is taken as \cite{Ref4,Ref5,Ref6,Ref7,Ref8,Ref9,Ref10,Ref11,Ref12,Ref13,Ref14,Ref15,Ref16,Ref17,Ref18,Ref19,Ref20,Ref21,Ref22,Ref23} 
\begin{equation}
{\mathcal L}= -V(\phi)F(X)
\end{equation}
where
\begin{equation}
 X={\frac{1}{2}{\partial_{\mu}{\phi}{\partial^{\mu}{\phi}}}}=\frac{1}{2}{\dot{\phi}^{2}}-\frac{1}{2}(\nabla\phi)^{2}
\end{equation}
and the pressure $P$ is given by $P={\mathcal L}=-V(\phi)F(X)$

Energy density $\rho$ for k-essence field is 
\begin{equation}
\rho=V(\phi)[F(X)-2X F_{X}]
\end{equation}
with 
$F_{X}=\frac{\partial F}{\partial X}$.\\
Considering constant potential ($V(\phi)$=constant, so that $\frac{\partial V}{\partial \phi}=0$) the conservation equation $(9)$ yields
\begin{eqnarray}
[2X F_{XX}+F_{X}]\dot{X}+4\frac{\dot{T}}{T}X F_{X}
\nonumber\\
+[2X F_{XX}
+F_{X}]\frac{1}{2}\frac{d}{dt}[(\nabla\phi)^{2}]=0
\end{eqnarray}
We now consider what types of fields $\phi(r,t)$ can lead to a scaling relation. Not all types of fields are amenable to a scaling relation.
It may be verified that with the decomposition $\phi(r,t)=\phi_{1}(r)\phi_{2}(t)$, no scaling relation can be obtained.
However, if the field permits a decomposition of the type $\phi(r,t)=\phi_{1}(r)+\phi_{2}(t)$, then a
scaling relation can be worked out. We shall  justify this {\it a posteriori}. As acceptable fields must permit Taylor expansions , so 
we may write around some point $(a,b)$,

$\phi(r,t) = \phi(a,b)+ \frac{\partial\phi (a,b)}{\partial r}(r-a) + \frac{\partial\phi (a,b)}{\partial t}(t-b) + higher~~ orders $

Here we have assumed that $(a,b)$ is not an extremum point of $\phi(r,t)$, so that the first derivatives exist at the point.
We also assume that higher order derivatives are vanishingly small. We work with this tangent plane expansion
of the fields. So the fields will be linear in $r$ and $t$. {\it Note that we do not require that $r$ and $t$ be small.} Further , the first term and the derivatives evaluated at the point $(a,b)$ are constants. Therefore , we may write  $\phi(r,t)= \phi(a,b)+ \phi_{1}(r)+\phi_{2}(t)$. The first (constant) term will not occur in the dynamics as 
its derivatives would vanish. So we may forget it and just write $\phi(r,t)=\phi_{1}(r)+\phi_{2}(t)$. {\it We shall show below in section 4 that solutions to $\phi_{1}(r)$ and $\phi_{2}(t)$ are indeed linear in $r,t$.}

With this decomposition of the fields, $(13)$ reduces to
\begin{equation}
[2X F_{XX}+F_{X}]\dot{X}+4\frac{\dot{T}}{T}X F_{X}=0
\end{equation}
Solving we obtain, 
\begin{equation}
X {F_{X}}^{2}=CT^{-4}
\end{equation}
 where $C$ is an integration constant. 

This is a scaling relation for inhomogeneous cosmology.  This is different from the scaling relation $ X {F_{X}}^{2}=Ca(t)^{-6}$\cite{Ref4,Ref5,Ref6,Ref7,Ref8,Ref9}, where $a(t)$ is a scale factor. 
Below we will show that with appropriate identifications, this is the 
same scaling relation present in FLRW homogeneous cosmology with 
dark energy lagrangian for k-essence cosmology\cite{Ref4,Ref5,Ref6,Ref7,Ref8,Ref9}.
\section{ Lagrangian}
\label{sec:2}
Equating k-essence energy density $(12)$ with $(7)$ we get
\begin{equation}
XF_{X}=\frac{1}{2}\left[F(X)-\frac{4}{3V(\phi)}\left(\frac{\dot{T}}{T}\right)^{2}+\frac{\Lambda}{V(\phi)}\right]
\end{equation}
and making use of scaling relation $(15)$ yields
\begin{equation}
F(X)=+\frac{4}{3V(\phi)}\left(\frac{\dot{T}}{T}\right)^{2}+2\sqrt{C}\sqrt{X}T^{-2}-\frac{\Lambda}{V(\phi)}
\end{equation}
Hence the k-essence lagrangian $(10)$ becomes
\begin{eqnarray}
&&{\mathcal L} =-\frac{4}{3}\left(\frac{\dot{T}}{T}\right)^{2}
-2V(\phi)\sqrt{C}\sqrt{X}T^{-2}+\Lambda 
\end{eqnarray}
Since we are considering constant potential, we will write $V(\phi)=V_{0}$ where $V_{0}$ is some constant and now let $q=\ln T$ and,  so $(18)$ becomes
\begin{equation}
{\mathcal L}=-c_{1}{\dot{q}}^{2}-c_{2}V_{0}\sqrt{X} e^{-2q}+\Lambda
\end{equation}
This is the $k$-essence lagrangian for inhomogeneous scalar field $\phi=\phi(r,t).$
\section{Euler Lagrange equation and its solution}
\label{sec:3}
There are two generalised co-ordinates in the lagrangian $(19)$, $q$ and $\phi$.\\
Euler Lagrange equation for $q$ yields
\begin{equation}
\frac{d}{dt}(\dot{q})=-\frac{c_{2}}{c_{1}}V_{0}\sqrt{X} e^{-2q}
\end{equation}
and Euler Lagrange equation for $\phi$ yields
\begin{equation}
\frac{d}{dt}[\frac{\partial \sqrt{X}}{\partial\dot{\phi}}e^{-2q}]=0
\end{equation}
First integral for equation $(21)$ yields
\begin{equation}
\frac{\partial \sqrt{X}}{\partial\dot{\phi}}e^{-2q}=\frac{1}{D}
\end{equation}
where $\frac{1}{D}$ is some constant.\\
 With $\phi(r,t)=\phi_{1}(r)+\phi_{2}(t)$ we have $X=\frac{1}{2}\dot{\phi_{2}}^{2}(t)-\frac{1}{2}(\nabla{\phi_{1}(r)})^{2}$, and equation $(22)$ yields\\ 
\begin{equation}
\frac{1}{2}\dot{\phi_{2}}^{2}(t)-\frac{D^2}{4}\dot{\phi_{2}}^{2}(t)e^{-4q}=\frac{1}{2}(\nabla{\phi_{1}(r)})^{2}
\end{equation}
Now in $(23)$ the left hand side is a function of time $t$ while the right hand side is a function of $r$ only. Therefore we can write
\begin{equation}
\frac{1}{2}\dot{\phi_{2}}^{2}(t)-\frac{D^2}{4}\dot{\phi_{2}}^{2}(t)e^{-4q}=\frac{1}{2}(\nabla{\phi_{1}(r)})^{2}=B
\end{equation}
where $B$ is a constant.
From $(24)$ we get 
\begin{equation}
\dot{\phi_{2}}(t)^{2}=\frac{2B}{(1-\frac{D^2}{2}e^{-4q})}
\end{equation}
and
\begin{equation}
\phi_{1}(r)=\sqrt{2B}r
\end{equation}
{\it Therefore, $\phi_{1}(r)$ is linear in $r$}.
Substituting $(25)$ in $(20)$ yields
\begin{equation}
\ddot{q}=-\frac{c_{2}}{c_{1}}\frac{D}{2}V_{0}\frac{\sqrt{2B}}{(1-\frac{D^2}{2}e^{-4q})^{\frac{1}{2}}}e^{-4q}
\end{equation}
Integrating once 
\begin{equation}
\dot{q}=\alpha\bigg[1-\frac{D^2}{2}e^{-4q}\bigg]^{\frac{1}{4}}
\end{equation}
where $\alpha=\sqrt{-\frac{c_{2}V_{0}\sqrt{2B}}{c_{1}D}}$ . Integration constant has been chosen to be zero. We will always take 
$\alpha$ to be real. (For example, if $c_{1}, c_{2}, D, V_{0}$ are all real then taking the negative square root of $2B$ will ensure that $\alpha$ is real etc.).\\ 
Solving $(28)$ gives
\begin{equation}
t=\frac{1}{2\alpha}[-\frac{1}{2}\ln\left(\frac{1-(1-\frac{D^2}{2T^{4}})^{\frac{1}{4}}}{1+(1-\frac{D^2}{2T^{4}})^{\frac{1}{4}}}\right)
-\tan^{-1}{(1-\frac{D^2}{2T^{4}})^{\frac{1}{4}}}-2\beta]
\end{equation}
Here $\beta$ is an integration constant.\\
This is the relationship between $t$ and $T$. Let us analyse this solution. 
Let $ y={(1-\frac{D^2}{2T^{4}})^{\frac{1}{4}}} < 1$. This is a very plausible assumption because this only implies that $D^{2} > 0$.
So for small $y$ , equation $(30)$ becomes 

$$t=\frac{1}{2\alpha}[\frac{1}{2}\ln\left(\frac{1+y}{1-y}\right)
-\tan^{-1}{y}-2\beta]\approx\frac{1}{2\alpha}[\frac{1}{2}\ln(1+y)(1-y)^{-1}
-\tan^{-1}{y}-2\beta]\approx \frac{1}{2\alpha}[\frac{1}{2}\ln(1+y)^{2}
-\tan^{-1}{y}-2\beta]$$
$$\approx\frac{1}{2\alpha}[y-\frac{y^2}{2}-(y-\frac{y^3}{3!})-2\beta]\approx -\frac{y^2}{4\alpha}-\frac{\beta}{\alpha}
\approx -\frac{y^2}{4\alpha}-\frac{\beta}{\alpha}= -\frac{1}{4\alpha}(1-\frac{D^2}{2T^{4}})^{\frac{1}{2}}-\frac{\beta}{\alpha}
\approx -\frac{1}{4\alpha}(1-\frac{D^2}{4T^{4}}) -\frac{\beta}{\alpha}=-\frac{1}{4\alpha}-\frac{\beta}{\alpha}+ \frac{D^2}{16\alpha T^{4}}$$
where cubic and higher terms have been ignored,and we have let $D^2\rightarrow 0+$. Next choosing the arbitrary constant $\beta= -\frac{1}{4}$
we get 
\begin{equation}
T(t)=\frac{D^{1/2}}{2{\alpha}^{1/4} t^{1/4}}
\end{equation}
With this , solution for $\phi_{2}(t)$ is 
\begin{equation}
\phi_{2}(t)=-\frac{\sqrt{2B}}{8\alpha}(1-8\alpha t)^{-1/2}\approx -\frac{\sqrt{2B}}{8\alpha}(1+4\alpha t)
\end{equation}
where we have assumed that $\alpha=\sqrt{-\frac{c_{2}V_{0}\sqrt{2B}}{c_{1}D}} << 1$. {\it Therefore, $\phi_{2}(t)$ is also linear in $t$ and we have 
thus justified the splitting of the field into a sum of pure spatial and temporal parts,i.e., $\phi(r,t) =  \phi_{1}(r)+ \phi_{2}(t)$.}

If we revert to the tangent plane expansion language and choose expansion about the point $(a,b)\equiv (0,0)$, the equations $(26)$ and $(31)$ imply

$\phi(0,0)=-\frac{\sqrt{2B}}{8\alpha}~~;~~\frac{\partial\phi(0,0)}{\partial r}= \sqrt{2B};~~ \frac{\partial\phi (0,0)}{\partial t}= -\sqrt{\frac{B}{2}}$
\section{FLRW metric}
\label{sec:4}
 The plane LTB metric $(1)$ reduces to the flat FLRW metric in the limit $Y(t,r)\rightarrow a(t)r$ and $Y'(t,r)\rightarrow a(t)$ \cite{Ref10} :
\begin{equation}
ds^{2}=-dt^{2}+a^{2}(t)[dr^{2}+r^2d{\theta}^{2}+r^2\sin^{2}\theta d{\phi}^{2}]
\end{equation}
Since $Y=f^{\frac{2}{3}}$  and $f(r,t)=R(r)T(t)$ we get
\begin{equation}
Y(r,t)=R^{\frac{2}{3}}(r)T^{\frac{2}{3}}(t)=a(t)r
\end{equation}
This yields 
\begin{equation}
R^{\frac{2}{3}}(r)=r
\end{equation}
and
\begin{equation}
a(t)=T^{\frac{2}{3}}(t)\equiv \bigg(\frac{\dot\phi_{2}^2 D^2}{2\dot\phi_{2}^2 - 4 B}\bigg)^{\frac{1}{6}} 
\end{equation}
where we have used equation $(25)$.
 
Thus we can say $T^{\frac{2}{3}}(t)$ plays the role of cosmological scale factor for inhomogeneous plane LTB metric. Further $(35)$ means $ T^{4} = a(t)^{6}$. 
So the scaling relation $(15)$ becomes the same as in a homogeneous 
FLRW cosmology, i.e.,  $XF_{X}^{2}=Ca(t)^{-6}$ \cite{Ref4,Ref5,Ref6,Ref7,Ref8,Ref9}. 

An interesting aspect of these solutions are that in the limit that the LTB metric reduces to the FLRW metric, the FLRW metric can be written completely in terms of the dark energy scalar fields as follows :
\begin{equation}
ds^2\nonumber\\
=dt^2-\bigg(\frac{\dot\phi_{2}^2 D^2}{2\dot\phi_{2}^2 - 4 B}\bigg)^{\frac{1}{6}}
2B [ d\phi_{1}^2 + \phi_{1}^2 d\theta^2 + \phi_{1}^2 sin^2\theta d\Phi^2]\nonumber\\
\end{equation}

\section{Deceleration parameter $q_{0}$}
\label{sec:5}
Deceleration parameter is given by
\begin{equation}
q_{0}=-\frac{\ddot{a}a}{{\dot{a}}^{2}}
\end{equation}
Using $T(t)=a^{\frac{3}{2}}(t)$ ,$q=\ln T$, and the solution $(30)$ 
\begin{equation}
q_{0}=-1-\frac{3}{8}= -1.37
\end{equation}

From the observation of Type 1a Supernovae (SNe 1a) by The Supernova Cosmology Project \cite{Ref33,Ref34} and the High-Z-Supernova search team \cite{Ref35,Ref36} it was first established that the universe is undergoing accelerated expansion. Recent observations of luminosity distances of the SNe 1a confirm this accelerated expansion. Other observations like Cosmic Microwave Background anisotropies measured with WMAP satellite \cite{Ref39} and Planck satellite \cite{Ref44,Ref45,Ref46,Ref47} also ensure an accelerated expansion of the universe . All these observations directly or indirectly suggest the negativity of the deceleration parameter $(q_{0})$ with values ranging from $q_{0}\approx -1$ \cite{Ref40} to $q_{0}\approx-0.53$ \cite{Ref41,Ref42,Ref43}.\\
Thus $(38)$ is consistent with the observational value for the late time ($t\rightarrow\infty$) acceleration of the universe. The interesting aspect is that this has been derived by developing a k-essence model for inhomogeneous cosmology \cite{Ref26,Ref27,Ref28,Ref29,Ref30,Ref31,Ref32}. This result is very encouraging as all other relevant k-essence model for late time acceleration of the universe were basically based on FLRW metric for  homogeneous cosmology.

\section{Equation of state parameter $\omega$}
\label{sec:6}
Equation of state parameter $\omega$ is given by 
\begin{equation}
\omega=\frac{P}{\rho}
\end{equation}
Hence from (7) and (8) we get
\begin{equation}
\omega=\frac{P}{\rho}=\frac{-\frac{4}{3}\frac{\ddot{T}}{T}+\Lambda}{\frac{4}{3}\left(\frac{\dot{T}}{T}\right)^{2}-\Lambda}
\end{equation}
Since $q=\ln T$ this becomes
\begin{equation}
\omega=\frac{-\frac{4}{3}(\ddot{q}+{\dot{q}}^{2})+\Lambda}{\frac{4}{3}{\dot{q}}^{2}-\Lambda}
\end{equation}
We are interested in late time cosmologies. Considering the solution  $(30)$ for $T$ and using  $q=ln T$ gives $\ddot{q}\rightarrow 0$ for 
$t\rightarrow\infty$. Therefore $(41)$ implies that the equation of state parameter  becomes
\begin{equation}
\omega = -1
\end{equation}
 This is consistent with the observational values, as all the observational data \cite{Ref35,Ref36,Ref37,Ref38,Ref39,Ref40,Ref41,Ref42,Ref43,Ref44,Ref45,Ref46,Ref47,Ref48,Ref49,Ref50} more or less establishes $\omega\approx-1$.\\

This also satisfies dark energy pressure condition which is negative i.e., negative pressure generates late time cosmic acceleration \cite{Ref4,Ref5,Ref6,Ref7,Ref8,Ref9,Ref10,Ref11,Ref12,Ref13,Ref14,Ref15,Ref16,Ref17,Ref18,Ref19,Ref20,Ref21,Ref22,Ref23,Ref24,
Ref25,Ref26,Ref27,Ref28,Ref29,Ref30,Ref31,Ref32,Ref33,Ref34,Ref35,Ref36,Ref37,Ref38,Ref39,Ref40,Ref41,Ref42,Ref43,Ref44,Ref45,Ref46,Ref47,Ref48,Ref49,Ref50}.

\section{Conclusion}
\label{sec:7}
 
We consider k-essence model for dark energy in an inhomogeneous universe characterised by the Lemaitre-Tolman- Bondi metric \cite{Ref1,Ref2,Ref3,Ref17}.  
A scaling relation for spherically symmetric k-essence scalar fields \cite{Ref13} 
$\phi(r,t)$ is obtained which reduces to  the known relation for  a homogeneous cosmology when the LTB metric reduces to the FLRW metric under certain identifications of the metric functions. A k-essence lagrangian is set up and the Euler-Lagrangian equations solved assuming $\phi(r,t)=\phi_{1}(r) + \phi_{2}(t)$. Solutions of the fields $\phi_{1}(r)$ and $\phi_{2}(t)$ are shown to exist that are linear in $r$ and $t$ respectively. The solutions enable the LTB metric functions to be related to the fields. The  LTB inhomogeneous universe exhibits late time  accelerated expansion i.e., cosmic acceleration driven by negative pressure.

The authors thank the honourable referees for enlightening suggestions to improve the manuscript.

\bibliographystyle{elsarticle-num} 

\begin{thebibliography}{00}

\bibitem{Ref1} A.Kraisinski, {Inhomogeneous Cosmological Models} (Cambridge University Press), (1997)

\bibitem{Ref2} K.Bolejko, M.N.Celerier, A.Krasinski,  Class.Quant.Grav,\textbf{28} (2011), Article 164002  

\bibitem{Ref3} L.P.Chimento, D.Pavon, Gen.Rel.Grav, \textbf{ 30} (1998), pp. 643-651 

 
\bibitem{Ref4} R.J.Scherrer, Phys.Rev.Lett,\textbf{ 93} (2004), Article 011301

\bibitem{Ref5} L.P.Chimento, Phys.Rev.D,\textbf{69}  (2004), Article 123517 

\bibitem{Ref6} L.P.Chimento, A.Feinstein, Modern Physics Letters A, \textbf{19} (2004), pp. 761-768 

\bibitem{Ref7} C.A-Picon, T.Damour, V.Mukhanov, Phys.Lett.B, \textbf{458} (1999), pp. 209-218 

\bibitem{Ref8} D.Gangopadhyay, S.Mukherjee, Phys.Lett.B, \textbf{665}(2008), pp. 121-124  

\bibitem{Ref9} D.Gangopadhyay, S.Mukherjee, Grav.Cosmol,\textbf{ 17} (2011), pp. 349-354  

\bibitem{Ref10} J.D-Santiago, J.L.C-Cota, Phys.Rev.D, \textbf{83} (2011), Article 063502 

\bibitem{Ref11} R.Putter, E.V.Linder, Astropart.Phys, \textbf{28} (2007), p. 263 

\bibitem{Ref12} F.Arroja, M.Sasaki, Phys.Rev.D, \textbf{81} (2010), Article 107301 

\bibitem{Ref13} R.Saitouand, S.Nojiri, Eur.Phys.J.C, \textbf{71} (2011)

\bibitem{Ref14} L.A.Garcia, J.M.Tejeiro, L.Castaneda, arXiv:1210.52259.

\bibitem{Ref15} C.A-Picon, T.Damour, V.Mukhanov, Phys.Lett.B, \textbf{458} (1999), p. 209 

\bibitem{Ref16} J.Garriga, V.Mukhanov, Phys.Lett.B, \textbf{458} (1999), p. 219 

\bibitem{Ref17} E.J.Copeland, M.Sami, S.Tsujikawa, Int.J.Mod.Phys.D, \textbf{15} (2006), pp. 1753-1936 

\bibitem{Ref18} L.P.Chimento, M.Forte, Phys.Rev.D, \textbf{73} (2006), Article 063502 

\bibitem{Ref19} K.Enqvist, Gen.Rel.Grav,\textbf{40} (2008), pp. 451-466

\bibitem{Ref20} C.H.Chuang,	Class.Quant.Grav, \textbf{ 25} (2008), Article 175001 

\bibitem{Ref21} M.N.Celerier, New Advances in Physics, \textbf{ 29} (2007), p.1  

\bibitem{Ref22} A. D. Rendall, Class.Quant.Grav, \textbf{23} (2006), pp. 1557-1570 	

\bibitem{Ref23} L. P. Chimento, A. S. Jakubi, D. Pavon, 	arXiv:gr-qc/9911030

\bibitem{Ref24} D. Wands, J. De-Santiago, Y. Wang, Class. Quantum Grav, \textbf{29} (2012), Article 145017 

\bibitem{Ref25} L. P. Chimento, A. S. Jakubi, D. Pavon, 	Phys.Rev.D, \textbf{60} (1999), Article 103501 

\bibitem{Ref26} K. Bolejko, M. Korzyński, Int. J. Mod. Phys.D, \textbf{26} (2017), Article 1730011 

\bibitem{Ref27} I.Bormotova, E.Kopteva, M.Churilova, Z.Stuchlik, Int.J.Mod.Phys.A, \textbf{35} (2020), Article 2040037

\bibitem{Ref28} K.Enqvist, Gen.Rel.Grav, \textbf{30} (2008), pp. 451-466

\bibitem{Ref29} C.H.Chuang, Class.Quant.Grav, \textbf{25} (2008), Article 175001

\bibitem{Ref30} R.J.Scherrer, Phys.Rev.D, \textbf{73} (2006), Article 043502

\bibitem{Ref31} Sh.Khosravi, E.Kourkchi, R.Mansouri, Y.Akrami, Gen.Rel.Grav, \textbf{40}  (2008), pp. 1047-1069, 

\bibitem{Ref32} R.A.Sussman, Class.Quant.Grav, \textbf{25} (2008), Article 015012

\bibitem{Ref33} S.Perlmutter, et al, Nature, \textbf{391} (1998), p. 51

\bibitem{Ref34} S.Perlmutter, et al, Astrophys.J, \textbf{517} (1999), p. 565

\bibitem{Ref35} B.P.Schimdt, et al, Astrophys.J, \textbf{507} (1998), p. 46

\bibitem{Ref36} N.W.Halverson, et al, Astrophys.J, \textbf{568} (2002), p. 38

\bibitem{Ref37} C.L.Bennett, et al, Astrophys.J.Suppl, \textbf{148} (2003), p. 1
 
\bibitem{Ref38} D.N.Spergel, et al, Astrophys.J.Suppl, \textbf{148} (2003), p. 175

\bibitem{Ref39} E.Komatsu, et al, Astrophys.J.Suppl, \textbf{192} (2011), p. 18

\bibitem{Ref40} A.G.Riess, et al, Astrophys.J, \textbf{116} (1998), pp. 1009-1038

\bibitem{Ref41} A.G.Riess, et al, Astrophys.J, \textbf{659} (2007), p. 98

\bibitem{Ref42} R.Giostri, M. Vargas dos Santos, I. Waga, et al. 2012, JCAP, 03, 027

\bibitem{Ref43} K.Kleidis, N.K. Spyrou , Astron.Astrophys., \textbf{A23} (2015), p.576

\bibitem{Ref44} P.A.R.Ade, et al, Astron.Astrophys, \textbf{A1} (2014), p. 571

\bibitem{Ref45} P.A.R.Ade, et al, Astron.Astrophys, \textbf{A12} (2014), p. 571

\bibitem{Ref46} P.A.R.Ade, et al, Astron.Astrophys, \textbf{A15} (2014), p. 571

\bibitem{Ref47} N.Aghanim, et al, Astron.Astrophys, \textbf{Special issue} (2020) 

\bibitem{Ref48} R.R.Caldwell, R.Dave, P.J.Steinhardt, Phys.Rev.Lett, \textbf{80} (1998), pp. 1582-1585

\bibitem{Ref49} M.Tegmark, et al, Phys.Rev.D, \textbf{69} (2004), Article 103501

\bibitem{Ref50} W.M.Wood-Vesey, et al, Astrophys.J, \textbf{666} (2007), pp. 694-715






 



\end{thebibliography}
{}

\end{document}